\documentclass[11pt,twoside,letterpaper]{article} 
\usepackage{times,fancyhdr}
\usepackage[dvips]{graphicx}
\usepackage{cite}
\usepackage{amsbsy}


\makeatletter 
\def\cleardoublepage{\clearpage\if@twoside \ifodd\c@page\else%
    \hbox{}%
    \thispagestyle{empty}%
    \newpage%
    \if@twocolumn\hbox{}\newpage\fi\fi\fi} 
\makeatother

\renewcommand{\thesection}{\arabic{section}.}
\renewcommand{\thesubsection}{\thesection\arabic{subsection}.}

\def\figurename{Figure}
\makeatletter
\renewcommand{\fnum@figure}[1]{\figurename~\thefigure.}
\makeatother

\def\tablename{Table}
\makeatletter
\renewcommand{\fnum@table}[1]{\tablename~\thetable.}
\makeatother

\begin{document}
\title{
{\begin{flushleft}
\vskip 0.45in
{\normalsize\bfseries\textit{Chapter~6}}
\end{flushleft}
\vskip 0.45in
\bfseries\scshape Rotation curves in Bose-Einstein Condensate Dark Matter Halos}}
\author{\bfseries\itshape M. Dwornik, Z. Keresztes, L. \'{A}. Gergely\thanks{E-mail address: gergely@physx.u-szeged.hu} \\
Department of Theoretical Physics, University of Szeged, \\
Tisza Lajos krt 84-86, Szeged 6720, Hungary \\
Department of Experimental Physics, University of Szeged, \\
D\'{o}m T\'{e}r 9, Szeged 6720, Hungary \\
}
\date{}
\maketitle
\thispagestyle{empty}
\setcounter{page}{1}

\section{ABSTRACT}

The study of the rotation curves of spiral galaxies reveals a nearly
constant cored density distribution of Cold Dark Matter. N-body simulations
however lead to a cuspy distribution on the galactic scale, with a central
peak. A Bose-Einstein condensate (BEC) of light particles naturally solves
this problem by predicting a repulsive force, obstructing the formation of
the peak. After succinctly presenting the BEC model, we test it against
rotation curve data for a set of 3 High Surface Brightness (HSB), 3 Low
Surface Brightness (LSB) and 3 dwarf galaxies. The BEC model gives a similar
fit to the Navarro-Frenk-White (NFW) dark matter model for all HSB and LSB
galaxies in the sample. For dark matter dominated dwarf galaxies the
addition of the BEC component improved more upon the purely baryonic fit
than the NFW component. Thus despite the sharp cut-off of the halo density,
the BEC dark matter candidate is consistent with the rotation curve data of all types of galaxies.

\section{INTRODUCTION}

Cosmological observations provide compelling evidence that about 95\% of the
content of the Universe resides in the unknown dark matter and dark energy
components \cite{planck1}. The former resides in bound systems as
non-luminous matter \cite{pers, salu}, the latter in the form of a
zero-point energy pervading the whole Universe. Dark matter is thought to be
composed of pressureless, cold, neutral, weakly interacting massive
particles, beyond those existing in the Standard Model of Particle Physics,
and not yet detected in accelerators or in dedicated direct and indirect
searches, excepting gravitational has been found. Therefore the possibility
that the Einstein (and Newtonian) theory of gravity breaks down at the scale
of galaxies cannot be excluded a priori. While some studies show that the
luminous matter alone can explain the rotation in the innermost galactic
regions \cite{sell, evans, palu}, dark matter is still required on the
larger scale. Several theoretical models, based on a modification of
Newton's law or of general relativity, have been proposed to explain the
behavior of the galactic rotation curves \cite{milg, sand, mof, mann,
roberts, boeh3, boeh1, berto, boeh2}. In brane world models, the galactic
rotation curves can be naturally explained without introducing dark matter 
\cite{mak, raha, gerg}. There is also a possibility that the rotation of
galaxies in the outermost regions could be driven by magnetic fields, rather
than dark matter \cite{battaner}.

In recent cosmological models the primordial density fluctuations are
generated during an inflationary period and they are the seeds of the
bottom-up structure formation model.\ The post-inflation regime is usually
described by the $\Lambda $CDM (cosmological constant + Cold Dark Matter)
model which is consistent with the vast majority of the available
observations, including the large scale matter distribution, the Ia type
supernovae observations and the temperature fluctuations in the cosmic
microwave background radiation \cite{peeb, padm, planck2}.

However, the investigation of spiral galaxies clearly shows that the mass
distribution of galactic-scale objects can not be explain satisfactory
within the framework of the $\Lambda $CDM cosmological model. The predicted
halo density profile is approximately isothermal over a large range in
radii, and it shows a well pronounced central cusp \cite{nav}. The
Navarro-Frenk-White (NFW)\ density profile is proportional to $1/r$ close to
the centre. On the observational side however, high-resolution rotation
curves show instead that the actual distribution of dark matter is much
shallower \cite{burkert}, presenting a constant density core. The Burkert
density profile shows a correlation between the enclosed surface densities
of luminous and dark matter in galaxies \cite{gent}. However the
astrophysical origin of this empirical density distribution remains
unaddressed.

The knowledge of the mass distribution of spiral galaxies is a crucial step
in the search for non-baryonic dark matter. Most of the present models
assume that, beside the stellar disk and bulge, there is a spherically
symmetric, massive dark matter halo, which dominates the total galaxy mass
and also determines the dynamics of the stellar disk at the outer regions.
Nevertheless Ref. \cite{jaloc} found that the mass distribution for some
galaxies cannot be spherical at larger radii and instead a flattened mass
distribution (global disk model) better approximates the gravitational
potential.

In this chapter we consider scalar field dark matter halos which have
undergone a Bose-Einstein condensation (BEC) \cite{sin, fuku, boeh1, chav1,
chav2, rind, harko1, harko2}. Below a critical temperature, bosons favor
joining highly populated low-energy states. At the end of this process,
bosons will occupy the same quantum ground state and form a coherent matter
wave, the Bose-Einstein condensate. In this state, the bosons exhibit a
repulsive interaction which prevents the formation of central density cusps
by gravitational attraction.

This chapter is organized as follows. The basic properties of the
Bose-Einstein condensed dark matter are reviewed in Section 2. Then the
theoretical predictions of the model are compared with the observed rotation
curve data of several types of galaxies (High Surface Brightness, Low
Surface Brightness and Dwarf Galaxies, respectively), in Section 3. We
discuss the results in Section 4.

\section{THE BOSE-EINSTEIN CONDENSATE}

An ideal Bose gas (a cloud of non-interacting bosons) confined in a box of
volume $V=L^{3}$ obeys the Bose-Einstein thermal distribution $f=\left\{
\exp \left[ \left( \epsilon -\mu \right) /k_{B}T\right] -1\right\} ^{-1}$,
where $\epsilon =p^{2}/2m$ stands for the energy of the bosons, determined
by their mass $m$ and magnitude $p$ of their 3-momentum vector, $\mu $ is
the chemical potential, $T$ the temperature and $k_{B}$ the Boltzmann
constant. The 3-momentum $\mathbf{p}=2\pi \hbar \mathbf{q}/L$ is discretized
through the dimensionless vector $\mathbf{q}$, with integer components. The
lowest energy state has $p=0$. The number of uncondensated bosons at
temperature $T$ is%
\begin{equation}
N_{T}=\sum\limits_{p\neq 0}\frac{1}{\exp \left[ \left( p^{2}/2m-\mu \right)
/k_{B}T\right] -1}~.  \label{NT}
\end{equation}%
If the energy levels follow each other densely, i.e. the the thermal energy
is much larger than the smallest energy spacing between the single-particle
levels:%
\begin{equation}
k_{B}T\gg \frac{2\pi ^{2}\hbar ^{2}}{mV^{2/3}}~,
\end{equation}%
the summation in Eq. (\ref{NT}) can be replaced by an integral over the
momentum ($\sum\limits_{\mathbf{p}}\rightarrow V\int d\mathbf{p}/\left( 2\pi
\hbar \right) ^{3}$) cf. Ref. \cite{pita}, thus:%
\begin{equation}
N_{T}=\frac{V}{\lambda _{T}^{3}}g_{3/2}\left[ \exp \left( \mu /k_{B}T\right) %
\right] ~.
\end{equation}%
Here 
\begin{equation}
\lambda _{T}=\hbar \sqrt{\frac{2\pi }{mk_{B}T}}~
\end{equation}%
is the thermal de Broglie wavelength and%
\begin{equation}
g_{3/2}\left( z\right) =\frac{2}{\sqrt{\pi }}\int_{0}^{\infty }dx\frac{\sqrt{%
x}}{e^{x}/z-1}
\end{equation}%
is a special Bose function, with $z=\exp \left( \mu /k_{B}T\right) $ and the
integral variable $x=p^{2}/2mk_{B}T$. At such temperatures when the ground
state is vacant, the value of the chemical potential can be expressed from
the relation $N_{T}=N$, with $N$ the total number of bosons. If the lowest
energy state $\epsilon _{0}=0$ is occupied, the chemical potential is given
by%
\begin{equation}
\mu =-k_{B}T\ln \left( 1+\frac{1}{n_{0}}\right) \approx -\frac{k_{B}T}{n_{0}}%
~,
\end{equation}%
with the ground state particle number $n_{0}$. In the thermodynamic limit $%
n_{0},N,V\rightarrow \infty $, $n=N/V,n_{0}/V\rightarrow $const, such that $%
n_{0}/N$ can be expressed as 
\begin{equation}
\frac{n_{0}}{N}=1-\left( \frac{T}{T_{c}}\right) ^{3/2}~\left( T\leq
T_{c}\right) ~.  \label{cond}
\end{equation}%
The critical temperature 
\begin{equation}
T_{c}=\frac{2\pi \hbar ^{2}}{mk_{B}}\left( \frac{n}{g_{3/2}\left( 0\right) }%
\right) ^{2/3}~,  \label{Tc}
\end{equation}%
with $g_{3/2}\left( 0\right) =2.612$, is typically low ($T_{c}=3.13$ K for $%
^{4}$He liquid at saturated vapour pressure), and represents the temperature
below which the bosons start to condensate into the lowest energy state. By
further decreasing the temperature, the relative number of particles in the
ground state increases. For $T>T_{c}$ the ground state is vacant.

The condition for the condensation $T<T_{c}$ can be rewritten as a relation
between the average distance of the bosons $l=\sqrt[3]{V/N}=n^{-1/3}$ and
their thermal de Broglie wavelength as 
\begin{equation}
l<\frac{\lambda _{T}}{\zeta ^{1/3}}\approx 0.73\lambda _{T}~.
\end{equation}%
With the bosons considered a quantum-mechanical wave packet of the order of
its de Broglie wavelength, the condensation occurs at the low temperatures
where their wavelengths overlap.

As the thermodynamic limit is never realized exactly, corrections arising
from the finite size slightly alter the value of the critical temperature,
for details see \cite{GrossmannHolthaus}, \cite{KetterleDruten}, \cite%
{KristenToms}, \cite{Haugerudetal}.

In a dense, non-ideal (self-interacting) Bose gas the particles can form
molecules, and they can reach a more stable state than a BEC. Two atoms can
form a molecule if a third particle takes momentum away. In a dilute gas
such a scenario can be avoided. Therefore BEC can be formed in a dilute and
ultracold Bose gases. The gas is considered dilute if the characteristic
length of the interaction $l_{int}$ is much smaller than the average
distance of the bosons, thus $l_{int}^{3}n\ll 1$. In a dilute gas the bosons
are weakly interacting through two-particle interactions. BEC can form in a
dilute, non-ideal Bose gas, however the condensate fraction is smaller and
the critical temperature is again altered \cite{Giorginietal}, \cite%
{Glaumetal} , \cite{Schutte}, \cite{dalfovo}. Experimentally, BEC has been
realized first by different groups in $^{87}$Rb (\cite{Anderson95}, \cite%
{Han98}, \cite{Ernst98}), and in $^{23}$Na (\cite{Davis95}, \cite{Hau98}),
in $^{7}$Li (\cite{Bradley95}).

\subsection{Mean field approximation, the Gross-Pitaevskii equation}

The static configuration of $N$ interacting scalar bosons placed in the
external potential $V_{ext}$ in a second quantized formalism is
characterized \cite{dalfovo} by the Hamiltonian operator 
\begin{eqnarray}
\hat{H} &=&\int d\mathbf{r}^{\prime }\hat{\Psi}^{+}\left( \mathbf{r}^{\prime
}\right) \left[ -\frac{\hbar ^{2}}{2m}\Delta ^{\prime }+V_{ext}\left( 
\mathbf{r}^{\prime }\right) \right] \hat{\Psi}\left( \mathbf{r}^{\prime
}\right)  \nonumber \\
&&+\frac{1}{2}\int d\mathbf{r}^{\prime }d\mathbf{r}^{\prime \prime }\hat{\Psi%
}^{+}\left( \mathbf{r}^{\prime }\right) \hat{\Psi}^{+}\left( \mathbf{r}%
^{\prime \prime }\right) V_{self}\left( \mathbf{r}^{\prime }-\mathbf{r}%
^{\prime \prime }\right) \hat{\Psi}\left( \mathbf{r}^{\prime }\right) \hat{%
\Psi}\left( \mathbf{r}^{\prime \prime }\right) ~.  \label{HamiltonOp}
\end{eqnarray}%
The operators (singled out by hats) are taken in the \textit{Schr\"{o}%
dringer picture}.\footnote{%
The Hamiltonian operator coincides in the Schr\"{o}dringer and Heisenberg
pictures since it does not depend explicitly on time.} The boson field
operators $\hat{\Psi}\left( \mathbf{r}^{\prime }\right) $ and $\hat{\Psi}%
^{+}\left( \mathbf{r}^{\prime }\right) $ annihilate and create a particle at
the position $\mathbf{r}^{\prime }$, while $\Delta ^{\prime }$ is the
3-dimensional Laplacian with respect to the coordinates $\mathbf{r}^{\prime
} $. The \textit{repulsive}, two-body interatomic potential is 
\begin{equation}
V_{self}=\lambda \delta \left( \mathbf{r}-\mathbf{r}^{\prime }\right)
\end{equation}%
with 
\begin{equation}
\lambda =\frac{4\pi \hbar ^{2}a}{m}
\end{equation}%
a self-coupling constant, given in terms of the scattering length $a$.

The field operator $\hat{\Psi}\left( \mathbf{r}\right) $ is decomposed in
terms of the single-particle annihilation operators $\hat{a}_{\alpha }$ as%
\begin{equation}
\hat{\Psi}\left( \mathbf{r}\right) =\sum\limits_{\alpha }\Psi _{\alpha
}\left( \mathbf{r}\right) \hat{a}_{\alpha }~,
\end{equation}%
where $\Psi _{\alpha }$ is the wave function of single-particle state $%
\left\vert \alpha \right\rangle $. The summation is taken over the
single-particle state. The functions $\Psi _{\alpha }$ are orthonormal and
form a complete set of single-particle wave functions, i.e.%
\begin{equation}
\sum\limits_{\alpha }\Psi _{\alpha }\left( \mathbf{r}\right) \Psi _{\alpha
}^{\ast }\left( \mathbf{r}^{\prime }\right) =\delta \left( \mathbf{r}-%
\mathbf{r}^{\prime }\right) ~,  \label{completness}
\end{equation}%
where a star denotes the complex conjugation. Denoting the particle numbers
in some state (labeled $\alpha $) by $n_{\alpha }$, the bosonic annihilation 
$\hat{a}_{\alpha }$ and creation $\hat{a}_{\alpha }^{+}$ operators act on
the Fock space as 
\begin{eqnarray}
\hat{a}_{\alpha }\left\vert n_{0},n_{1},...,n_{\alpha },...\right\rangle  &=&%
\sqrt{n_{\alpha }}\left\vert n_{0},n_{1},...,n_{\alpha }-1,...\right\rangle
~, \\
\hat{a}_{\alpha }^{+}\left\vert n_{0},n_{1},...,n_{\alpha },...\right\rangle
&=&\sqrt{n_{\alpha }+1}\left\vert n_{0},n_{1},...,n_{\alpha
}+1,...\right\rangle ~,
\end{eqnarray}%
and satisfy the following commutation relations:%
\begin{equation}
\left[ \hat{a}_{\alpha },\hat{a}_{\beta }^{+}\right] =\delta _{\alpha \beta
}~,~\left[ \hat{a}_{\alpha },\hat{a}_{\beta }\right] =0~,~\left[ \hat{a}%
_{\alpha }^{+},\hat{a}_{\beta }^{+}\right] =0~.  \label{comm1}
\end{equation}%
The numbers $n_{\alpha }$ are the eigenvalues of the operator $n_{\alpha }=%
\hat{a}_{\alpha }^{+}\hat{a}_{\alpha }$. The commutation relations 
\begin{equation}
\left[ \hat{\Psi}\left( \mathbf{r}\right) ,\hat{\Psi}^{+}\left( \mathbf{r}%
^{\prime }\right) \right] =\delta \left( \mathbf{r}-\mathbf{r}^{\prime
}\right) ~,~\left[ \hat{\Psi}\left( \mathbf{r}\right) ,\hat{\Psi}\left( 
\mathbf{r}^{\prime }\right) \right] =0~,~\left[ \hat{\Psi}^{+}\left( \mathbf{%
r}\right) ,\hat{\Psi}^{+}\left( \mathbf{r}^{\prime }\right) \right] =0
\end{equation}%
for the field operators follow from (\ref{comm1}) and (\ref{completness}).

Since the Hamiltonian operator is time-independent, the boson field operator 
$\hat{\Psi}\left( \mathbf{r},t\right) $ in the \textit{Heisenberg picture} is%
\begin{eqnarray}
\hat{\Psi}\left( \mathbf{r},t\right)  &=&e^{i\hat{H}t/\hbar }\hat{\Psi}%
\left( \mathbf{r}\right) e^{-i\hat{H}t/\hbar }=\sum\limits_{\alpha }\Psi
_{\alpha }\left( \mathbf{r}\right) e^{i\hat{H}t/\hbar }\hat{a}_{\alpha }e^{-i%
\hat{H}t/\hbar }  \nonumber \\
&=&\sum\limits_{\alpha }\Psi _{\alpha }\left( \mathbf{r}\right) \hat{a}%
_{\alpha }\left( t\right) ~,
\end{eqnarray}%
with%
\begin{equation}
\hat{a}_{\alpha }\left( t\right) =e^{i\hat{H}t/\hbar }\hat{a}_{\alpha }e^{-i%
\hat{H}t/\hbar }~.
\end{equation}%
Similar equations are valid for the respective adjoint operators $\hat{\Psi}%
^{+}\left( \mathbf{r},t\right) $, $\hat{a}_{\alpha }^{+}\left( t\right) $.
The field operators obeys the Heisenberg equation, i.e.:%
\begin{eqnarray}
i\hbar \frac{\partial }{\partial t}\hat{\Psi}\left( \mathbf{r},t\right)  &=&%
\left[ \hat{\Psi}\left( \mathbf{r},t\right) ,\hat{H}\right]   \nonumber \\
&=&\left[ -\frac{\hbar ^{2}}{2m}\Delta +V_{ext}\left( \mathbf{r}\right)
+\lambda \hat{\Psi}^{+}\left( \mathbf{r},t\right) \hat{\Psi}\left( \mathbf{r}%
,t\right) \right] \hat{\Psi}\left( \mathbf{r},t\right) ~.  \label{Heisenberg}
\end{eqnarray}%
Instead Eq. (\ref{Heisenberg}), it is effective computationally to use a 
\textit{mean-field approximation}. The basic idea \cite{Bogoliubov1947} is
to separate the BEC contribution in the field operator as 
\begin{equation}
\hat{\Psi}\left( \mathbf{r},t\right) =\Psi _{0}\left( \mathbf{r}\right) \hat{%
a}_{0}\left( t\right) +\hat{\Psi}^{\prime }\left( \mathbf{r},t\right) ~,
\end{equation}%
where the zero subscript denotes the ground-state and $\hat{\Psi}^{\prime
}\left( \mathbf{r},t\right) $ carries the effects of the excited states. BEC
occurs when the number of particles $n_{0}\left( t\right) $ in the
condensate becomes very large, hence the states with $n_{0}\left( t\right) $
and $n_{0}\left( t\right) +1$ correspond to the same configuration. In this
case $\hat{a}_{0}\left( t\right) \approx \hat{a}_{0}^{+}\left( t\right)
\approx \sqrt{n_{0}\left( t\right) }$ and the expectation value of the BEC
contribution is 
\begin{equation}
\psi (\mathbf{r},t)=\sqrt{n_{0}\left( t\right) }\Psi _{0}\left( \mathbf{r}%
\right) ~,
\end{equation}%
the wave function of the condensate. By comparison the contribution from the
non-condensed part is small, therefore $\hat{\Psi}^{\prime }\left( \mathbf{r}%
,t\right) $ represents a perturbation with a negligible expectation value in
the leading order approximation.

The probability density 
\begin{equation}
\rho \left( \mathbf{r},t\right) =\left\vert \psi (\mathbf{r},t)\right\vert
^{2}~,
\end{equation}%
by the choice $\int d\mathbf{r}\left\vert \Psi _{0}\left( \mathbf{r}\right)
\right\vert ^{2}=1$ is normalized to 
\begin{equation}
n_{0}\left( t\right) =\int d\mathbf{r}\rho \left( \mathbf{r},t\right) ~,
\label{norm}
\end{equation}%
such that $\rho \left( \mathbf{r},t\right) $ also represents the number
density of the condensate.

In the leading order approximation (neglecting the contribution of the
excited states) Eq. (\ref{Heisenberg}) becomes 
\begin{equation}
i\hbar \frac{\partial }{\partial t}\psi (\mathbf{r},t)=\left[ -\frac{\hbar
^{2}}{2m}\Delta +V_{ext}\left( \mathbf{r}\right) +\lambda \rho \left( 
\mathbf{r},t\right) \right] \psi (\mathbf{r},t)~,  \label{Heisenberg2}
\end{equation}%
known as the Gross-Pitaevskii equation \cite{Gross1}, \cite{Gross2}, \cite%
{Pitaevskii} which describes the Bose-Einstein condensate in the mean-field
approximation.

\subsection{Madelung hydrodynamic equations}

In order to find a solution of Eq. (\ref{Heisenberg2}) it is worth to use
the Madelung representation of complex wave-functions \cite{Madelung}, \cite%
{Sonego}: 
\begin{equation}
\psi (\mathbf{r},t)=\sqrt{\rho \left( \mathbf{r},t\right) }\exp \left[ \frac{%
i}{\hbar }S\left( \mathbf{r},t\right) \right] ~,
\end{equation}%
where the real-valued phase $S\left( \mathbf{r},t\right) $ has the dimension
of an action. The real part of Eq. (\ref{Heisenberg2}) gives 
\begin{equation}
\frac{\partial S}{\partial t}+\frac{1}{2m}\left( \nabla S\right)
^{2}+\lambda \rho +V_{ext}+V_{Q}=0~,  \label{HJ}
\end{equation}%
corresponding to a generalized Hamilton-Jacobi equations with quantum
correction potential 
\begin{equation}
V_{Q}=-\frac{\hbar ^{2}}{2m}\frac{\Delta \sqrt{\rho }}{\sqrt{\rho }}~.
\label{VQ}
\end{equation}%
The imaginary part of (\ref{Heisenberg2}) becomes a continuity equation 
\begin{equation}
\frac{\partial \rho }{\partial t}+\nabla \left( \rho \mathbf{v}\right) =0~,
\label{cont}
\end{equation}%
while the gradient of (\ref{HJ}) gives 
\begin{equation}
m\rho \left[ \frac{\partial \mathbf{v}}{\partial t}+\left( \mathbf{v\cdot }%
\nabla \right) \mathbf{v}\right] =-\nabla p-\rho \nabla V_{ext}-\rho \nabla
V_{Q}~.  \label{Euler}
\end{equation}%
Here we have introduced the notations 
\begin{equation}
\mathbf{v}=\frac{\nabla S\left( \mathbf{r}\right) }{m}~,
\end{equation}%
and 
\begin{equation}
p=\frac{\lambda }{2}\rho ^{2}~.
\end{equation}%
The $i$th component of the last term of Eq. (\ref{Euler}) can be rewritten
as \cite{Barceloetal}%
\begin{equation}
\rho \nabla _{i}V_{Q}=\sum_{j}\nabla _{j}\sigma _{ij}^{Q}~,
\end{equation}%
where%
\begin{equation}
\sigma _{ij}^{Q}=-\frac{\hbar ^{2}}{4m}\rho \nabla _{i}\nabla _{j}\ln \rho ~.
\end{equation}%
Eqs. (\ref{cont}) and (\ref{Euler}) correspond to the usual continuity and
Euler equations of fluid mechanics, with $\mathbf{v}$ the classical velocity
field, $p$ the pressure and $\sigma _{ij}^{Q}$ representing a quantum
correction to the stress tensor. Eqs. (\ref{cont}) and (\ref{Euler}) are the
Madelung hydrodynamic equations.

\subsection{Self-gravitating, spherically symmetric BEC distribution in the
Thomas-Fermi approximation}

A stationary state $\psi $ is 
\begin{equation}
\psi (\mathbf{r},t)=\sqrt{\rho \left( \mathbf{r}\right) }\exp \left( \frac{%
i\mu }{\hbar }t\right) ~,
\end{equation}%
with $\mu =$const. Then the continuity equation is automatically satisfied
while Eq. (\ref{HJ}) leads to 
\begin{equation}
V_{ext}+V_{Q}+\lambda \rho =\mu ~.  \label{fieldEq}
\end{equation}

The quantum correction potential $V_{Q}$ has significant contribution only
close to the bound \cite{Wang}, therefore it can be neglected as compared to
the self-interaction term $\lambda \rho $. This \textit{Thomas-Fermi
approximation} becomes increasingly accurate with an increasing number of
particles \cite{Liebetal}.

If $V_{ext}\left( \mathbf{r}\right) /m$ is the Newtonian gravitational
potential created by the condensate, it satisfies the Poisson equation: 
\begin{equation}
\Delta \frac{V_{ext}}{m}=4\pi G\rho _{BEC}~,  \label{Poisson}
\end{equation}%
where $\rho _{BEC}=m\rho $ is the mass density of the BEC and $G$ is the
gravitational constant.

The Laplacian of Eq. (\ref{fieldEq}) and Eq. (\ref{Poisson}) give 
\begin{equation}
\Delta \rho _{BEC}+\frac{4\pi Gm^{2}}{\lambda }\rho _{BEC}=0~.
\label{Heisenberg4}
\end{equation}%
For a spherical symmetric distribution this simplifies to 
\begin{equation}
\frac{d^{2}\left( r\rho _{BEC}\right) }{dr^{2}}+\frac{4\pi Gm^{2}}{\lambda }%
\left( r\rho _{BEC}\right) =0~,
\end{equation}%
with the solution \cite{Wang}, \cite{boeh1}: 
\begin{equation}
\rho _{BEC}\left( r\right) =\rho _{BEC}^{\left( c\right) }\frac{\sin kr}{kr}%
~,  \label{massdensity}
\end{equation}%
where%
\begin{equation}
k=\sqrt{\frac{Gm^{3}}{\hbar ^{2}a}}~
\end{equation}%
and $\rho _{BEC}^{\left( c\right) }\equiv \rho _{BEC}\left( 0\right) $ is a
central density, determined from the normalization condition (\ref%
{Heisenberg}) as 
\begin{equation}
\rho _{BEC}^{\left( c\right) }=\frac{n_{0}mk^{3}}{4\pi ^{2}}~.
\end{equation}%
The exists of a central finite density is exactly the required feature which
represents an advantage over cuspy dark matter profiles derived from N-body
simulation.

At the end of this subsection we comment on the validity of the Thomas-Fermi
approximation. The quantum correction potential (\ref{VQ}), rewritten in
spherical coordinates and inserted into Eq. (\ref{fieldEq}) generates
constant and $\rho _{BEC}^{-2}$ terms, while the contribution of the
self-interaction term is proportional to $\rho _{BEC}$. Close to the
boundary $R_{BEC}$ therefore the Thomas-Fermi approximation fails.
Multiplying Eq. (\ref{fieldEq}) by $\left( \rho _{BEC}/\rho _{BEC}^{\left(
c\right) }\right) ^{2}$, with $\rho _{BEC}$ given by Eq. (\ref{massdensity}%
), and integrating on the range $r\in \left[ 0,R_{BEC}\right] $ gives the
global weight of these terms as $\hbar ^{2}/km$ for the $V_{Q}$ term and $%
\hbar ^{2}an_{0}$ for the self-interaction term, respectively. Therefore the
Thomas-Fermi approximation holds valid for $n_{0}\gg 1/ka$, a condition also
obtained by a different method in Ref. \cite{Wang}.

\subsection{BEC dark matter halo}

In the following, we investigate the possibility that dark matter halos are
Bose-Einstein condensates\textbf{\ }\cite{sin}.

The size of the BEC galactic dark matter halo is defined by $\rho
(R_{BEC})=0 $, giving $k=\pi /R_{BEC}$, i.e. 
\begin{equation}
R_{BEC}=\pi \sqrt{\frac{\hbar ^{2}a}{Gm^{3}}}~.
\end{equation}%
The mass profile of the BEC\ halo is then given by 
\begin{eqnarray}
m_{BEC}(r) &=&4\pi \int_{0}^{r}\rho _{BEC}(r)r^{2}dr  \nonumber \\
&=&\frac{4\pi \rho _{BEC}^{(c)}}{k^{2}}r\left( \frac{\sin kr}{kr}-\cos
kr\right) ~.
\end{eqnarray}%
The contribution of the BEC halo to the velocity profile of the particles
moving on circular orbit under Newtonian gravitational force is 
\begin{equation}
v^{2}\left( r\right) =\frac{4\pi G\rho _{BEC}^{(c)}}{k^{2}}\left( \frac{\sin
kr}{kr}-\cos kr\right) ~.  \label{vel}
\end{equation}%
This has to be added to the respective baryonic contribution.

\section{CONFRONTING THE BEC MODEL WITH ROTATION CURVE DATA}

In order to test the validity of the BEC dark matter model, we confront the
rotation curve data of a sample of 3 High Surface Brightness (HSB) galaxies,
3 Low Surface Brightness (LSB) galaxies and 3 dwarf galaxies, with both the
NFW dark matter and the BEC density profiles.

The commonly used NFW model is based on the numerical simulations of
dark-matter halos in the $\Lambda $CDM framework \cite{nav}. The mass
density profile is given by

\begin{equation}
\rho _{NFW}(r)=\frac{\rho _{s}}{\left( r/r_{s}\right) \left(
1+r/r_{s}\right) ^{2}}~,
\end{equation}%
where there are two fit parameters $\rho _{s}$ and $r_{s}$.

The mass within a sphere with radius $r=yr_{s}$ is then given by 
\begin{equation}
M(r)=4\pi \rho _{s}r_{s}^{3}\left[ \ln (1+y)-\frac{y}{1+y}\right] ~,
\end{equation}%
where $y$ is a dimensionless radial coordinate.

\subsection{HIGH SURFACE BRIGHTNESS GALAXIES}

\begin{figure*}[tbp]
\begin{center}
\resizebox{13.9cm}{!} {
\begin{tabular}{ccc}
\includegraphics[height=5cm, angle=360]{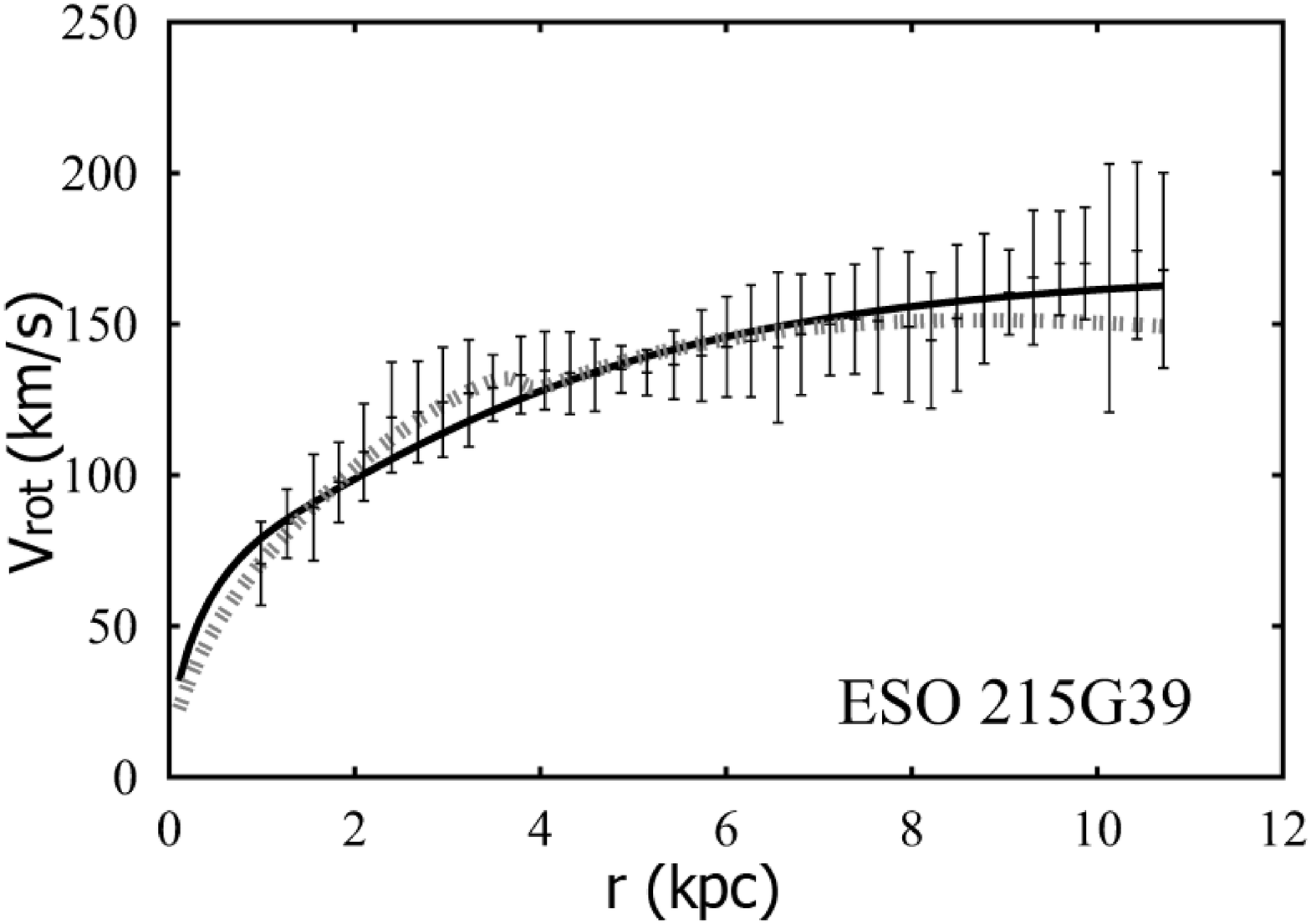} &
\includegraphics[height=5cm, angle=360]{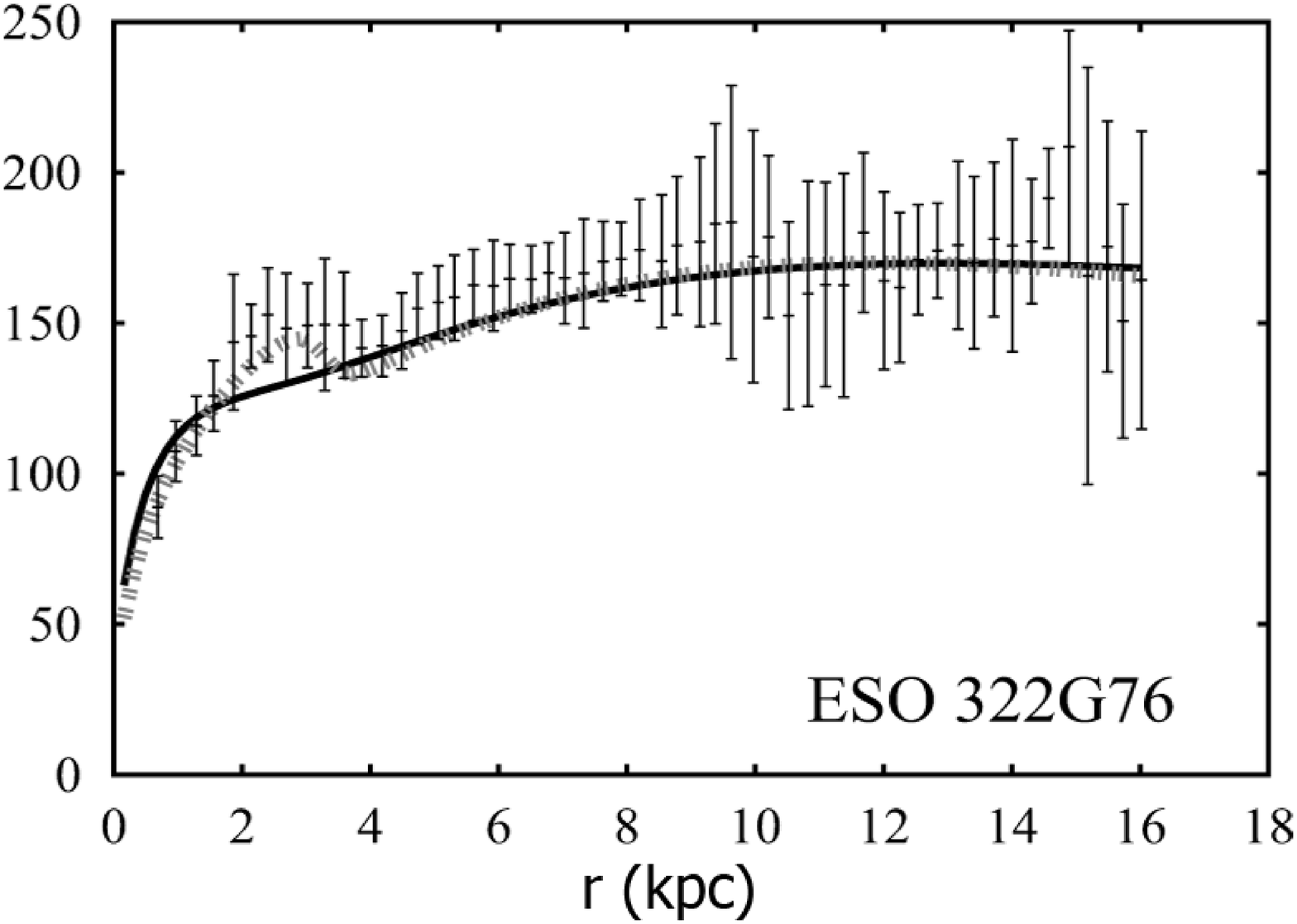} &
\includegraphics[height=5cm, angle=360]{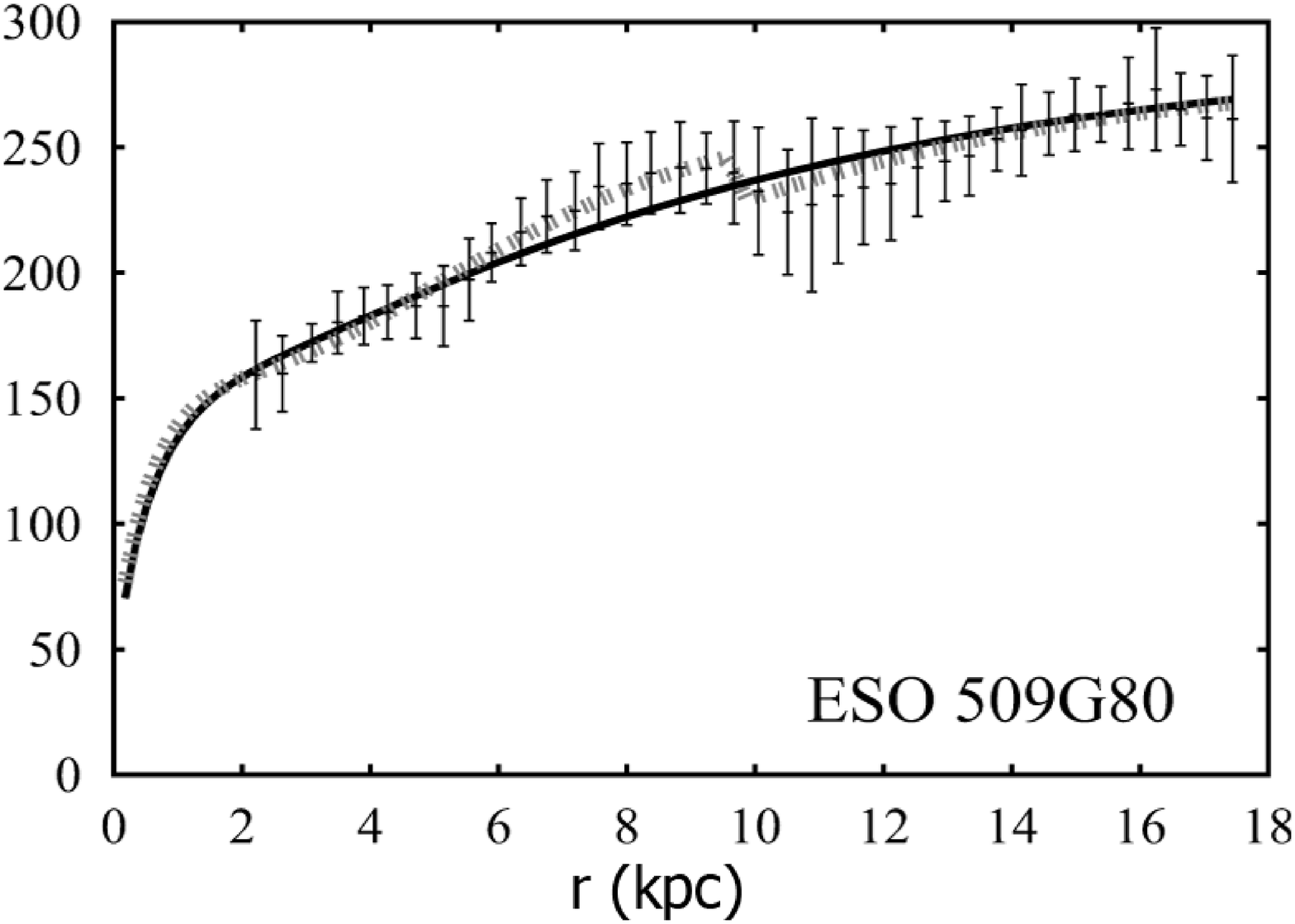} \\
&  &
\end{tabular}
}
\end{center}
\caption{Best fit curves for the HSB galaxy sample. The dashed lines hold in
the baryonic matter + BEC model, while the solid lines in the baryonic
matter + NFW model. In the second and third example the curves run very
close to each other.}
\label{hsbrot}
\end{figure*}

We follow the method described in Ref. \cite{gerg}. In a High Surface
Brightness galaxy we decompose the baryonic component into a thin stellar
disk and a spherically symmetric bulge. We assume that the mass distribution
of the bulge component follows the deprojected luminosity distribution, the
proportionality factor being the mass-to-light ratio. We estimate the bulge
parameters from a S\'{e}rsic $r^{1/n}$ bulge model, fitted to the optical
I-band galaxy light profiles.

\subsubsection{Bulge contribution}

The surface brightness profile of the spheroidal bulge component of each
galaxy is described by a generalized S\'{e}rsic function \cite{ser} 
\begin{equation}
I_{b}(r)=I_{0,b}\exp \left[ -\left( \frac{r}{r_{0}}\right) ^{1/n}\right] ~,
\label{Ib}
\end{equation}%
where $I_{0,b}$ is the central surface brightness of the bulge, $r_{0}$ is
its characteristic radius and $n$ is the shape parameter of the
magnitude-radius curve.

The respective mass over luminosity is the mass-to-light ratio, for the Sun
being $\gamma _{\odot }=5133$ kg W$^{-1}$. In what follows, the
mass-to-light ratio of the bulge $\sigma $ will be given in units of $\gamma
_{\odot }$ (solar units), while the masses in units of the solar mass $%
M_{\odot }=1.98892\times 10^{30}$ kg. The radial distribution of visible
mass is given by the radial distribution of light obtained from the
bulge-disk decomposition. Thus the mass of the bulge within the projected
radius $r$ is proportional to the surface brightness encompassed by this
radius: 
\[
M_{b}(r)=\sigma \frac{\mathcal{N}(D)}{F_{\odot }}2\pi
\int\limits_{0}^{r}I_{b}(r)rdr,
\]%
where $F_{\odot }\left( D\right) $ is the apparent flux density of the Sun
at a distance $D$ Mpc, $F_{\odot }\left( D\right) =2.635\times
10^{6-0.4f_{\odot }}\;\mathrm{mJy}{\ }$, with $f_{\odot }=4.08+5\lg \left(
D/1\;\mathrm{Mpc}\right) +25~\mathrm{mag}$, and 
\begin{equation}
\mathcal{N}(D)=4.4684\times 10^{-35}D^{-2}\;\mathrm{m}^{-2}\;\mathrm{arcsec}%
^{2}.
\end{equation}%
Therefore the contribution of the bulge to the rotational velocity is 
\begin{equation}
v_{b}^{2}(r)=\frac{GM_{b}(r)}{r}~.
\end{equation}%
\begin{table*}[tbp]
\begin{center}
\resizebox{13.9cm}{!} {
\begin{tabular}{c|c|c|c|c|c|c|c}
Galaxy & $D$ & $I_{0,b}$ & $n$ & $r_{0}$ & $r_{b}$ & $I_{0,d}^{HSB}$ & $
h^{HSB}$ \\ \hline
& Mpc & $\mathrm{mJy/arcsec} ^2$ &  & kpc & kpc & $\mathrm{mJy/arcsec} ^2$ & 
kpc \\ \hline\hline
ESO215G39 & 61.29 & 0.1171 & 0.6609 & 0.78 & 2.58 & 0.0339 & 4.11 \\ 
ESO322G76 & 64.28 & 0.2383 & 0.8344 & 0.91 & 4.50 & 0.0251 & 5.28 \\ 
ESO509G80 & 92.86 & 0.2090 & 0.7621 & 1.10 & 4.69 & 0.0176 & 11.03 \\ 
&  &  &  &  &  &  & 
\end{tabular}
}
\end{center}
\caption{The distances ($D$) and the photometric parameters of the 3 HSB
galaxy sample. Bulge parameters: the central surface brightness ($I_{0,b}$),
the shape parameter ($n$), the characteristic radius ($r_{0}$) and radius of
the bulge ($r_{b}$). Disk parameters: central surface brightness ($%
I_{0,d}^{HSB}$) and length scale ($h^{HSB}$) of the disk.}
\label{Tablehsbphoto}
\end{table*}

\subsubsection{Disk contribution}

In a spiral galaxy, the radial surface brightness profile of the disk
exponentially decreases with the radius \cite{free} 
\begin{equation}
I_{d}(r)=I_{0,d}^{HSB}\exp \left( -\frac{r}{h^{HSB}}\right) ,  \label{Id}
\end{equation}%
where $I_{0,d}^{HSB}$ is the disk central surface brightness and $h^{HSB}$
is a characteristic disk length scale. The contribution of the disk to the
circular velocity is \cite{free} 
\begin{equation}
v_{d}^{2}(x)=\frac{GM_{D}^{HSB}}{2h^{HSB}}x^{2}(I_{0}K_{0}-I_{1}K_{1}),
\end{equation}%
where $I_{n}$ and $K_{n}$ are the modified Bessel functions calculated at $%
x/2=r/2h^{HSB}$ and $M_{D}^{HSB}$ is the total mass of the disk.

\subsubsection{Confrontation with rotation curve data}

Therefore the rotational velocity in a HSB galaxy receives the following
contributions 
\[
v_{tg}^{2}(x)=v_{b}^{2}(x)+v_{d}^{2}(x)+v_{DM}^{2}(x)~. 
\]

We confront the BEC+baryonic model with ($HI$ and $H_{\alpha }$) rotation
curve data of 3 galaxies already employed in Ref. \cite{gerg}, which were
extracted from a larger sample given in Ref. \cite{palu} by requiring i)
sufficient and accurate data for each galaxy and ii) manifest spherical
structure of the bulge (no visible rings and bars). For comparison, the
NFW+baryonic model is also tested on the same sample, by plotting the
respective rotation curves in both models. The results are represented on
Fig. \ref{hsbrot}.

For the HSB galaxy sample we have derived the best fitting values of the
baryonic model parameters $I_{0,b}$, $n$, $r_{0}$, $r_{b}$, $I_{0,d}^{HSB}$
and $h^{HSB}$ from the available photometric data. The BEC and NFW
parameters were calculated (together with the corresponding baryonic
parameters) by fitting these models to the rotation curve data. These are
collected in the tables~\ref{Tablehsbphoto} and \ref{Tablehsbpar}.

\begin{table*}[tbp]
\begin{center}
\resizebox{13.9cm}{!} {
\begin{tabular}{c|c|c|c|c|c|c|c|c|c|c|c}
Galaxy & $\sigma $(BEC) & $M_{D}^{HSB}$(BEC) & $R_{BEC}$ & $\rho^{(c)}_{BEC}$ & $\chi _{\min }^{2}$ (BEC)&
$\sigma $(NFW)& $M_{D}^{HSB}$(NFW) & $r_{s}$ & $\rho_{s}$ & $\chi _{\min }^{2}$(NFW) &1$\sigma$ \\ \hline
 & $\odot$ & $10^{10} M_{\odot} $ & $kpc$ & $10^{-21}kg/m^3$ &   & $\odot$ & $10^{10} M_{\odot}$ & $kpc$ & $10^{-24}kg/m^3$ & \\ \hline\hline
ESO215G39 & 0.7 & 4.96 & 50 & 0.2 & 22.9 & 0.6 & 3.84 & 187 & 14.7 & 22.22 & 34.18 \\
ESO322G76 & 0.8 & 9.08 & 5 & 0.5 & 48.31 & 0.9 & 8.29 & 920 & 0.7 & 49.02 & 53.15  \\
ESO509G80 & 1.3 & 52.02 & 7 & 1.6 & 19.77 & 0.9 & 11 & 22 & 800 & 33.48 & 36.3\\
\end{tabular}
}
\end{center}
\caption{The best fit parameters and the minimum values ($\protect\chi %
_{\min }^{2}$) of the $\protect\chi ^{2}$ statistics for the 3 HSB galaxies.
Columns 2-5 give the BEC model parameters (radius $R_{BEC}$ and central
density $\protect\rho _{BEC}^{(c)}$ of the BEC halo) and the corresponding
baryonic parameters (mass-to-light ratio $\protect\sigma \left( BEC\right) $
of the bulge and total mass of the disk $M_{D}^{HSB}\left( BEC\right) $).
Columns 7-10 give the NFW model parameters (scale radius $r_{s}$ and
characteristic density $\protect\rho _{s}$ of the halo) and the
corresponding baryonic parameters (mass-to-light ratio $\protect\sigma %
\left( NFW\right) $ of the bulge and total mass of the disk $%
M_{D}^{HSB}\left( NFW\right) $). The 1$\protect\sigma $ confidence levels
are shown in the last column (these are the same for both models). For all
galaxies $\protect\chi _{\min }^{2}$ are within the 1$\protect\sigma $
confidence level. Both model fittings give similar $\protect\chi _{\min
}^{2} $ values.}
\label{Tablehsbpar}
\end{table*}

\subsection{LOW SURFACE BRIGHTNESS GALAXIES}

Low Surface Brightness (LSB) galaxies are characterized by a central surface
brightness at least one magnitude fainter than the night sky. These galaxies
form the most unevolved class of galaxies \cite{imp} and have low star
formation rates as compared to their HSB counterparts \cite{mcg}. LSB
galaxies show a wide spread of colours ranging from red to blue \cite{neil}
and represent a large variety of properties and morphologies. Although the
most commonly observed LSB galaxies are dwarfs, a significant fraction of
LSB galaxies are large spirals \cite{beij}.

Our model LSB galaxy consists of a thin stellar+gas disk and a cold dark
matter component in a form of BEC. The disk component is the same as at the
HSB galaxies, the surface brightness profile is \cite{free} 
\[
I_{d}(r)=I_{0,d}^{LSB}\exp \left( -\frac{r}{h^{LSB}}\right) ~, 
\]
where $I_{0,d}^{LSB}$ is the central surface brightness and $h^{LSB}$ is the
disk length scale. We can calculate the disk contribution to the circular
velocity as 
\begin{equation}
v_{d}^{2}(r)=\frac{GM_{D}^{LSB}}{2h^{LSB}}x^{2}(I_{0}K_{0}-I_{1}K_{1}),
\end{equation}
similarly to the case of HSB galaxies.

Therefore for a generic projected radius $r$, the rotational velocity in
this combined model can be written as 
\[
v_{tg}^{2}(r)=v_{d}^{2}(r)+v_{DM}^{2}~. 
\]

\begin{figure*}[tbp]
\begin{center}
\resizebox{13.9cm}{!} {
\begin{tabular}{ccc}
\includegraphics[height=5cm, angle=360]{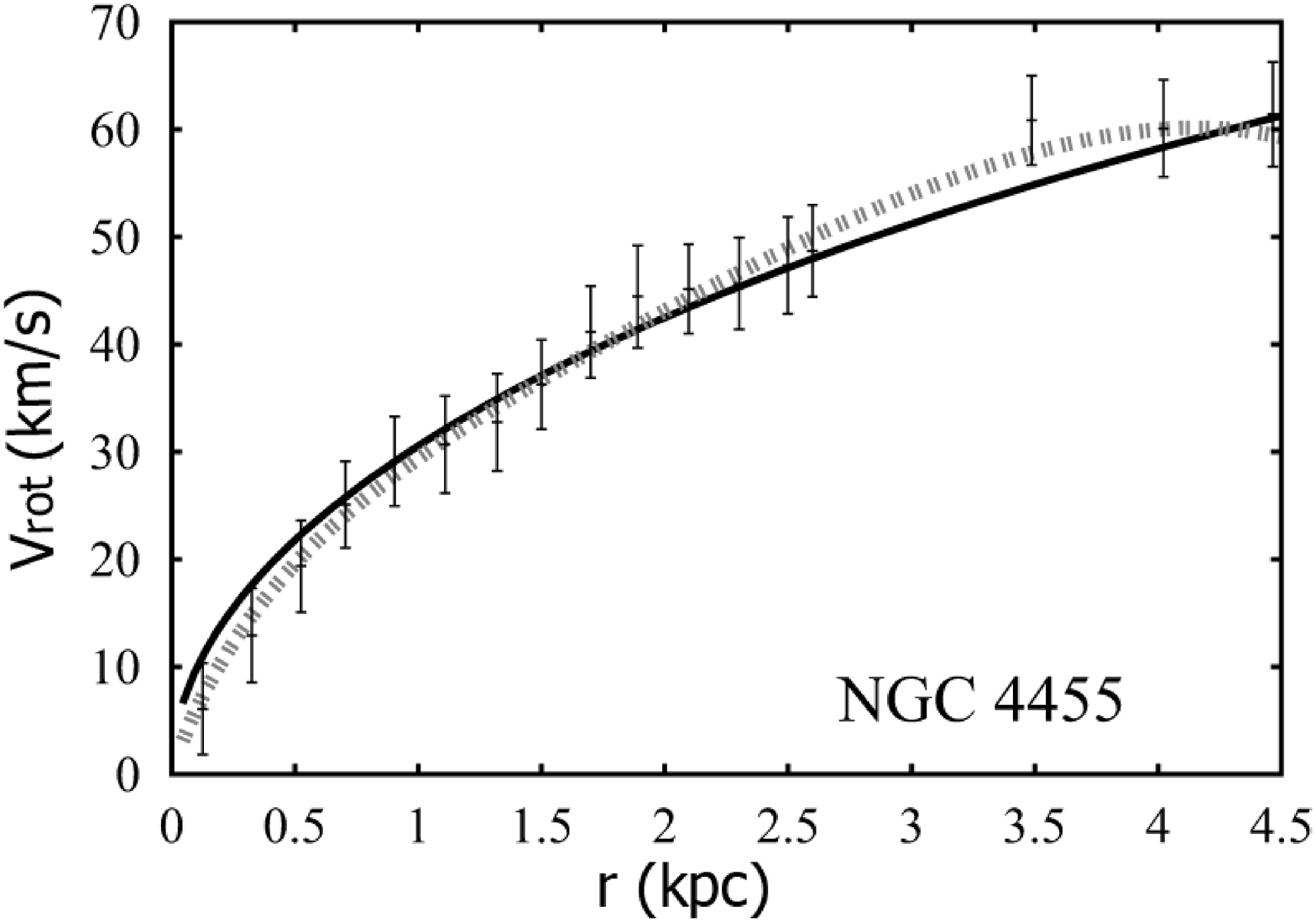} & 
\includegraphics[height=5cm, angle=360]{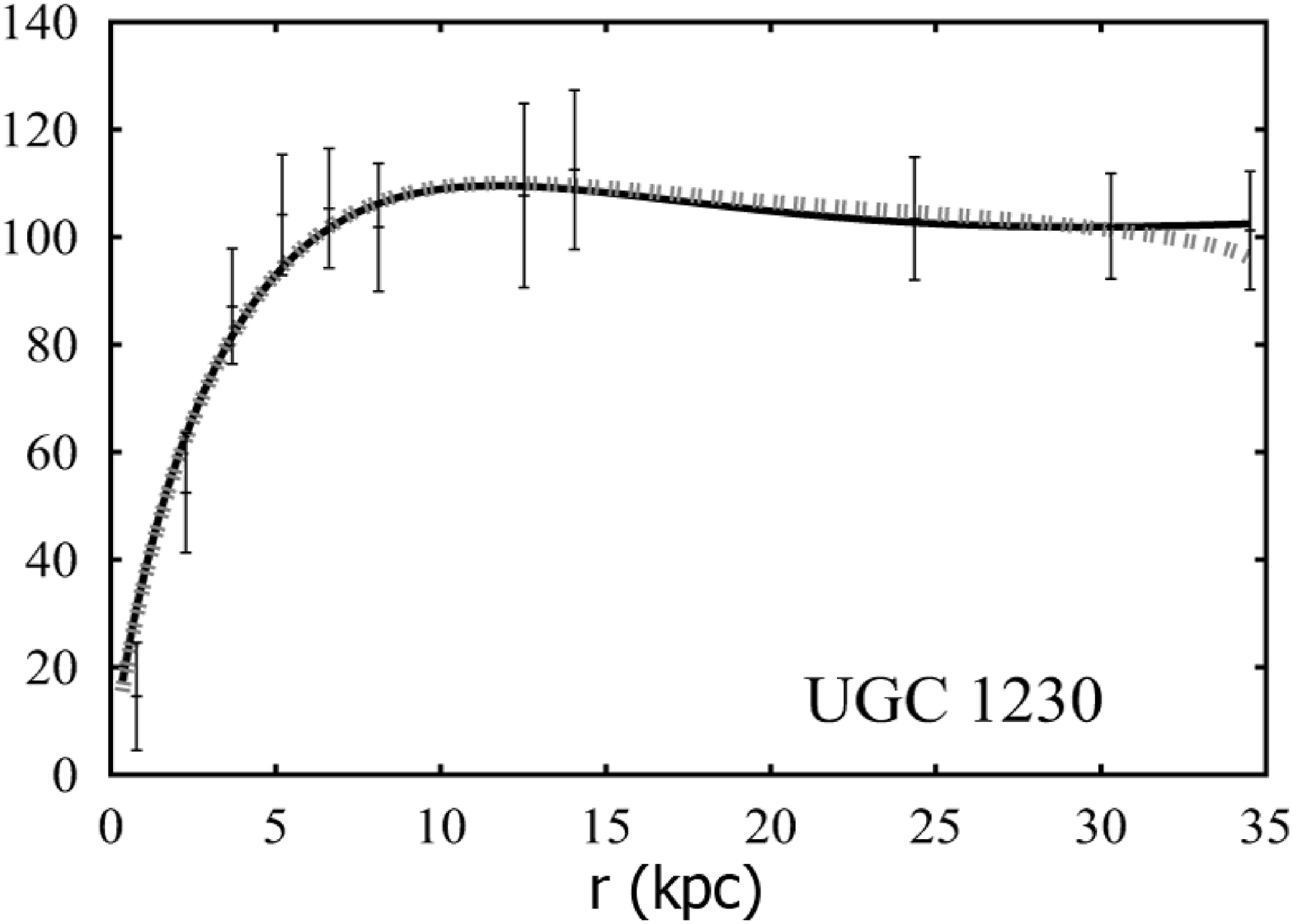} & 
\includegraphics[height=5cm, angle=360]{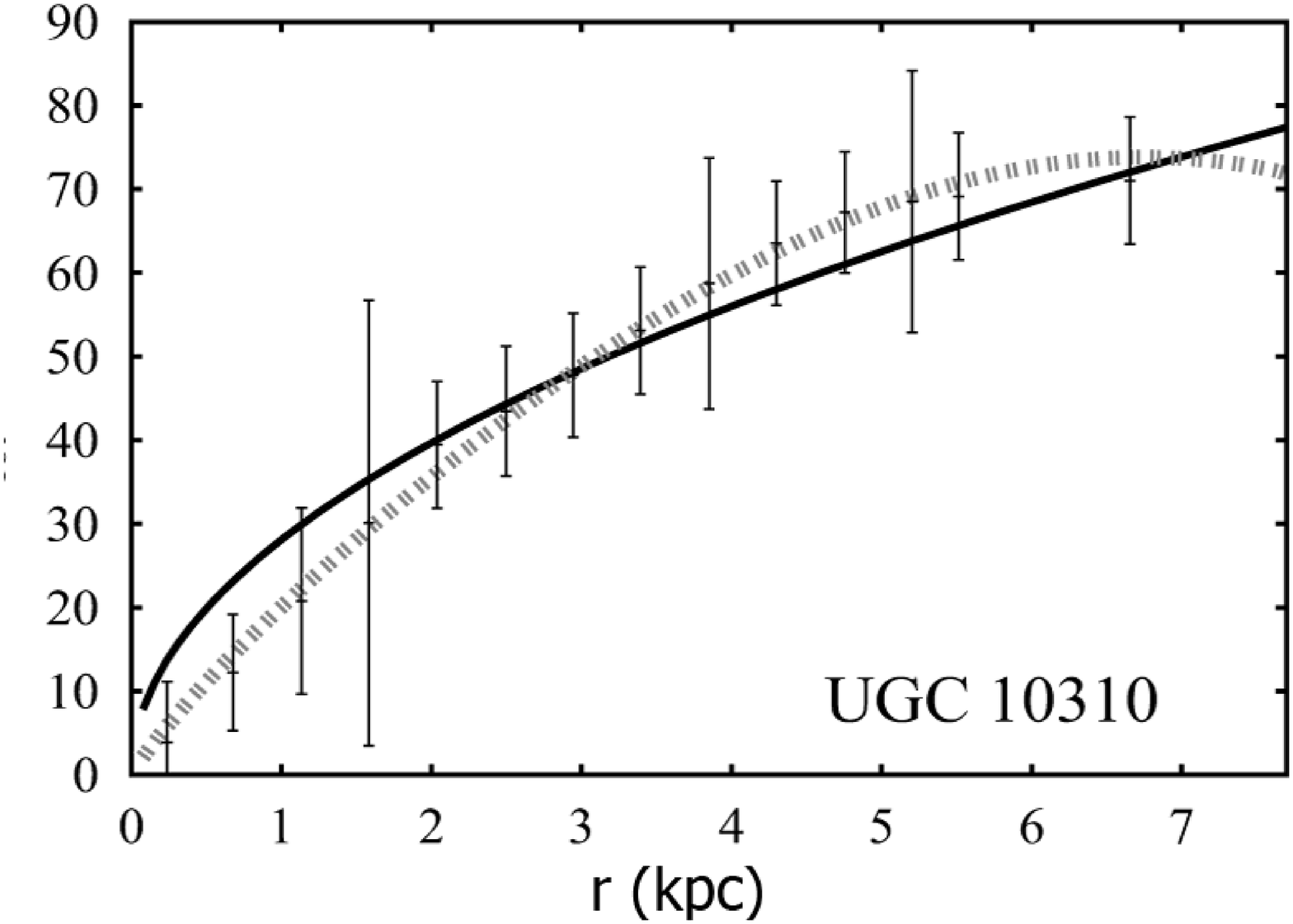} \\ 
&  & 
\end{tabular}
}
\end{center}
\caption{Best fit curves for the 3 LSB galaxies (the dashed lines show the
combined BEC+baryonic model, the solid lines show the combined NFW+baryonic
profiles).}
\label{rotcurvLSB}
\end{figure*}

We follow the analysis of the BEC model with the rotation curves of 3 LSB
galaxies taken from a larger sample \cite{blok}. These high quality rotation
curve data are based on both $HI$ and $H\alpha $ measurements. From a $\chi
^{2}$-test we have determined the model parameters in both the BEC+baryonic
and NFW+baryonic models, these are shown in Table \ref{Tablelsb}. The fitted
curves are represented on Fig.~\ref{rotcurvLSB}.

\begin{table*}[tbp]
\begin{center}
\resizebox{13.9cm}{!} {
\begin{tabular}{c|c|c|c|c|c|c|c|c|c|c|c}
Galaxy & $D$ & $h^{LSB}$ & $M_{D}^{LSB}(BEC) $ & $\rho^{(c)}_{BEC} $ & $R_{BEC}$ & $\chi _{\min }^{2}(BEC)$
& $M_{D}^{LSB}(NFW) $ & $\rho_{s}$ & $r_{s}$ & $\chi _{\min }^{2}(NFW)$ & 1$\sigma$ \\ \hline
& $Mpc$ & $kpc$ & $10^{9} M_{\odot}$ & $10^{-21}kg/m^3$ & $kpc$ & & $10^{9} M_{\odot}$ & $10^{-24}kg/m^3$ & $kpc$ & & \\ \hline\hline
NGC 4455 & 6.8 & 0.7  & 0.236 & 1.4& 5.6 & 9.4 & 0 & 62 & 39 & 30.37 & 18.11\\
UGC 1230 & 51 & 4.5  & 27 & 0.1& 27.7 & 19.86 & 26.2 & 2 & 291 & 19.84 & 8.17\\ 
UGC 10310 & 15.6 & 1.9  & 0.412 & 1 & 7.7 & 2.68 & 0 & 3 & 660 & 28.78 & 13.74\\

\end{tabular}
}
\end{center}
\caption{The best fit BEC and NFW parameters of the 3 LSB galaxies. }
\label{Tablelsb}
\end{table*}

\subsection{DWARF GALAXIES}

Dwarf galaxies are the most common galaxies in the observable Universe.
About 85\% of the known galaxies in the Local Volume are dwarfs \cite{karac}%
. Dwarf galaxies are defined as galaxies having an absolute magnitude
fainter than $M_{B}\sim -16$ mag, and more extended than globular clusters 
\cite{tamm}.

The formation history of dwarf galaxies is not well-understood. According to 
\cite{wu}, dwarf galaxies formed at the centers of subhalos orbiting within
the halos of giant galaxies. Five main classes of dwarf galaxies are
distinguished based on their optical appearance: dwarf ellipticals, dwarf
irregulars, dwarf spheroidals, blue compact dwarfs, and dwarf spirals. The
last one type can be regarded as the very small end of spirals \cite{matt}.

All dwarf galaxies have central velocity dispersions $\sim 6\div 25$ km/s 
\cite{mate}. If the systems are in dynamic equilibrium, the mass derived
from these velocity dispersions is much larger than the derived stellar
mass. Therefore the dwarf galaxies are among the darkest objects ever
observed in the Universe, hence they play an important role in the study of
dark matter distribution on small scales. As dwarf galaxies are supposed to
be dark matter dominated at all radii, they are ideal objects to prove or
falsify various alternative gravity theories \cite{capo}.

To test the BEC model, we have selected a sample of 3 dwarf galaxies for
which high resolution rotation curves are available. We performed the
rotation curve fitting with the BEC+baryonic and the NFW+baryonic models,
respectively. The baryonic components were the same as in the case of LSB
galaxies. However, the length scales of the stellar disks are not available
for this sample, therefore they are calculated from $\chi ^{2}$
minimization, too. This comparison allows us to test the viability of our
model.

\begin{figure*}[tbp]
\begin{center}
\resizebox{13.9cm}{!} {
\begin{tabular}{ccc}
\includegraphics[height=5cm, angle=360]{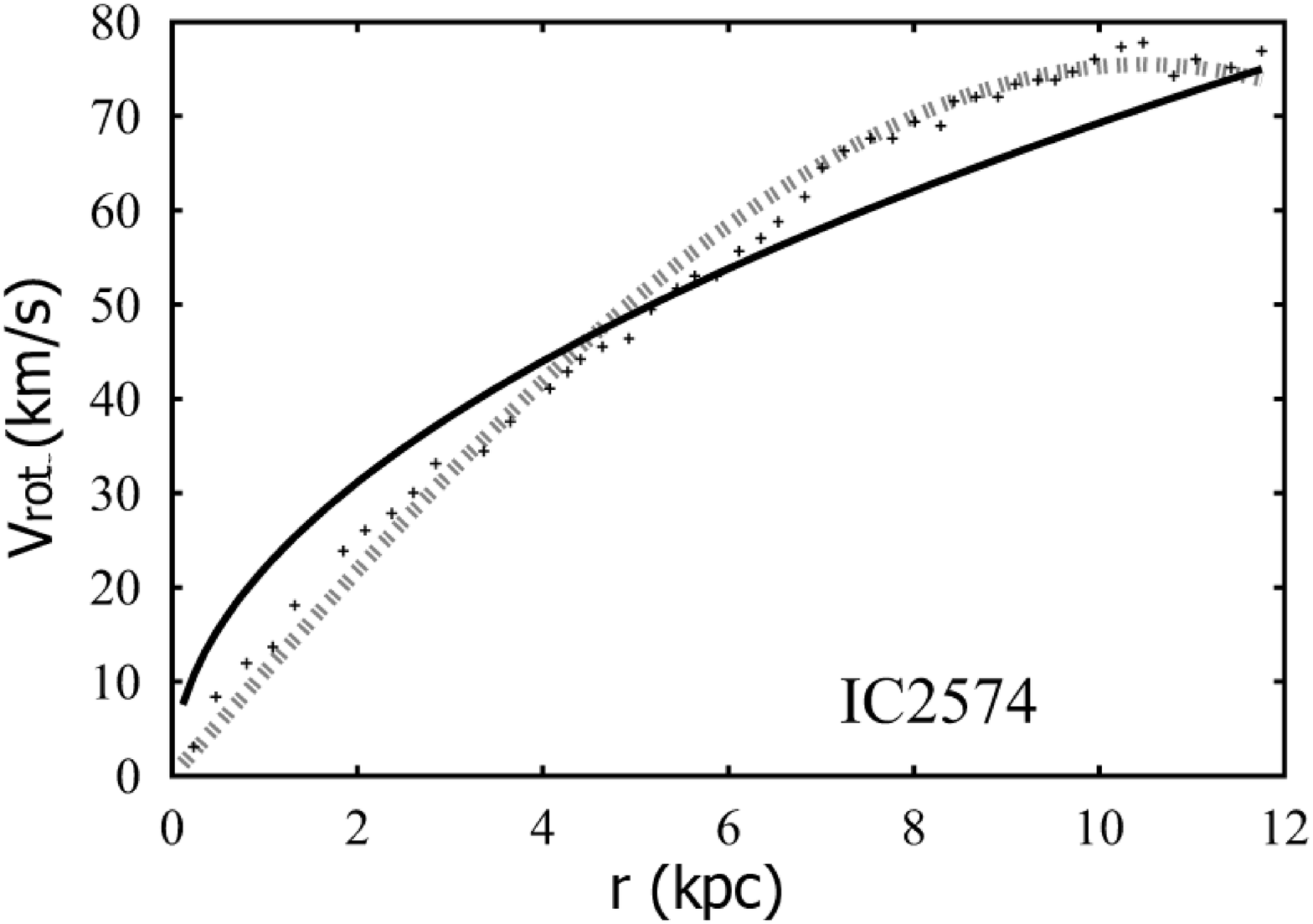} & 
\includegraphics[height=5cm, angle=360]{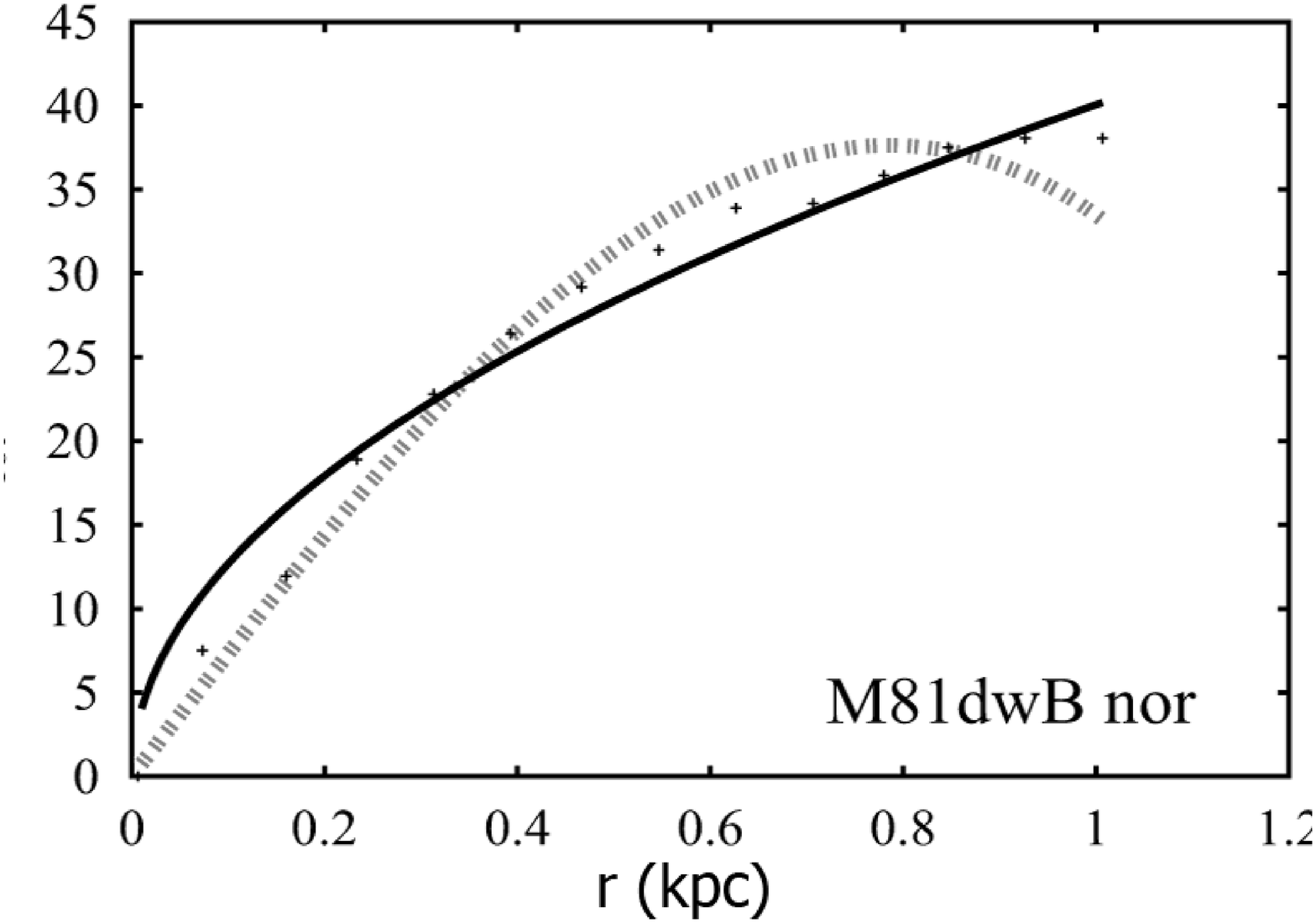} & 
\includegraphics[height=5cm, angle=360]{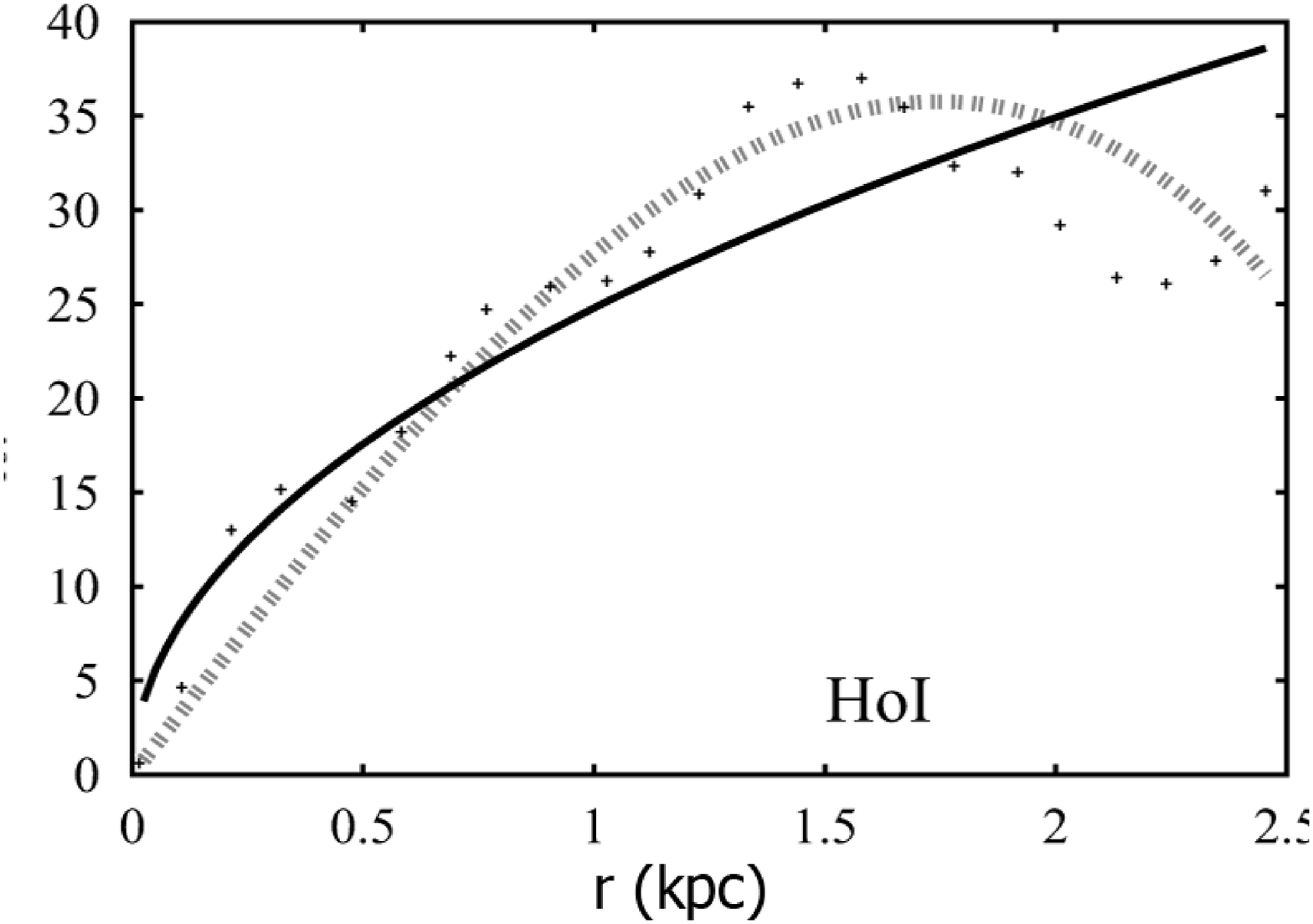} \\ 
&  & 
\end{tabular}
}
\end{center}
\caption{The best fit curves for the dwarf galaxy sample. The BEC+baryonic
model (dashed curves) gives a better fit in all cases then the NFW+baryonic
model (solid lines). In both cases the fit was performed with the same
baryonic model.}
\label{dwarfcombined}
\end{figure*}

For the investigated dwarf galaxies the best fit BEC and NFW parameters are
shown in Table~\ref{Tabledwarfbec} and the fitted rotation curves are
represented in Fig.~\ref{dwarfcombined}.

\begin{table*}[tbp]
\begin{center}
\resizebox{13.9cm}{!} {
\begin{tabular}{c|c|c|c|c|c|c|c|c|c|c|c}
Galaxy & $h^{dwarf}(BEC)$ & $M_{D}^{dwarf}(BEC) $ & $\rho^{(c)}_{BEC} $ & $R_{BEC}$ & $\chi _{\min }^{2}(BEC)$
& $h^{dwarf}(NFW)$ & $M_{D}^{dwarf}(NFW) $ & $\rho_{s}$ & $r_{s}$ & $\chi _{\min }^{2}(NFW)$ & 1$\sigma$ \\ \hline
& $kpc$ & $10^{9} M_{\odot}$ & $10^{-21}kg/m^3$ & $kpc$ & & $kpc$& $10^{9} M_{\odot}$ & $10^{-24}kg/m^3$ & $kpc$ & & \\ \hline\hline
IC 2574 & 1.2 & 0.1122 & 0.4 & 13 & 68.47 & 7.9 & 28.44 & 0 & 0 & 714.73 & 44.74  \\
HoI & 0.2 & 0.0107  & 3.6 & 1.9 & 95.26 & 0.9 & 0.533 & 0 & 0 & 241.30 & 20.27 \\
M81dwB nor & 0.9 & 1.023  & 3.7 & 0.7 & 6.19 & 0.7 & 0.705 & 0 & 0 & 8.4 & 10.42 \\
\end{tabular}
}
\end{center}
\caption{The best fit BEC+baryonic and NFW+baryonic parameters for the dwarf
galaxy sample.}
\label{Tabledwarfbec}
\end{table*}

\section{CONCLUSION}

We presented here a comprehensive introduction into the theory of
Bose-Einstein condensates (BEC), with emphasis on a spherically symmetric
self-gravitating BEC in the Thomas-Fermi approximation.

In the BEC model, large-scale structures like clusters or superclusters of
galaxies form similarly as in the Cold Dark Matter model with cosmological
constant, thus all predictions of the standard model at large scales are
well reproduced \cite{magana}. Moreover the BEC model can explain the
collisions of galaxy clusters \cite{lee} and the acoustic peaks of the
cosmic microwave background \cite{rodr}.

Such a BEC has been proposed recently as a galactic dark matter model and
has been partially tested by confronting with rotation curve data in Refs. 
\cite{harko1, robl, dwornik1}.

Here we have performed a much thorough test of this BEC dark matter model by
confronting with a sample of 3 HSB, 3 LSB and 3 dwarf galaxies and also
comparing the model predictions with those of the widely accepted NFW dark
matter model. We incorporated in all cases realistic baryonic models, taking
into account the particularities of the respective galaxy types. Beside the
rotation curve data for the HSB galaxies, the surface photometry data was
also available. The galaxies have different luminosities, disk length-scales
and surface brightnesses. Most of the rotation curve data were densely
distributed and uniform in quality.

We have also fitted the rotation curves of a sample of 3 \textit{dwarf
galaxies} with both the BEC+baryonic and NFW+baryonic dark matter models.
Since dwarf galaxies are supposed to be dark matter dominated, they provide
the strongest test of the compared models. The results are shown in Fig. \ref%
{dwarfcombined}. In all cases, the BEC dark matter model gave better results
than the NFW dark matter model.

For all galaxy types we have determined the BEC parameters $\rho
_{BEC},~R_{BEC}$, given in Tables \ref{Tablehsbpar}, \ref{Tablelsb}, and \ref
{Tabledwarfbec}.

For the investigated galaxies, we decomposed the circular velocity into its
baryonic and dark matter contributions: $%
v_{model}^{2}(r)=v_{baryonic}^{2}+v_{DM}^{2}$. The dark matter contribution
to the velocity is given by Eq.~(\ref{vel}). Then the rotation curves are $%
\chi ^{2}$ best-fitted with the baryonic parameters and the parameters of
the two dark matter halo models (BEC and NFW).

For the sample of \textit{HSB galaxies} we found a remarkably good agreement
of both dark matter models with the observations. The quality of the fits of
the BEC and NFW models with the rotation curve data was comparable.

For the sample of \textit{LSB galaxies}, the BEC model gave a slightly
better fit than the NFW model. We additionally found that adding the
baryonic component results in a better fit than the one presented in \cite%
{robl} for the pure BEC model.

We note that due to the sharp cut-off of the BEC halos the very distant behaviour of the
universal rotation curves (URCs) \cite{pers} is not expected to be reproduced, hence a 
modification of the BEC model on large distances would be desirable.

One suggestion takes into account the effects of the finite dark matter
temperature on the properties of the dark matter halos. However it turned
out, that these effects do not play a significant role in the description of
the dark matter halo density profiles \cite{harko3}. A possible solution is
including vortex lattices in the halo \cite{zinn, pita}. When a BEC is
rotated at a rate exceeding some critical frequency, quantized vortices can
be formed. This vortex lattice can influence the galactic rotation curve and
provide a flat velocity profile with oscillatory structure \cite{yu, robl}.

We conclude that the BEC dark matter model is well supported by rotation
curve data and it certainly deserves further attention.

\thispagestyle{fancy} \fancyhead{} 
\fancyhead[L]{In: Book Title \\ 
Editor: Editor Name, pp. {\thepage-\pageref{lastpage-01}}} 
\fancyhead[R]{ISBN 0000000000  \\
\copyright~2007 Nova Science Publishers, Inc.} \fancyfoot{} %
\renewcommand{\headrulewidth}{0pt} 

\vspace{2in}

\noindent \textbf{PACS} 05.45-a, 52.35.Mw, 96.50.Fm. \vspace{.08in}
\noindent \textbf{Keywords:} Dark matter halos.

\pagestyle{fancy} \fancyhead{} \fancyhead[EC]{Gergely et al.} %
\fancyhead[EL,OR]{\thepage} 
\fancyhead[OC]{Rotation curves in Bose-Einstein
Condensate Dark Matter Halos} \fancyfoot{} \renewcommand\headrulewidth{0.5pt}

\label{lastpage-01}

\end{document}